\documentclass[preprint, aps, superscriptaddress]{revtex4-1}
\usepackage[amssymb]{SIunits}
\usepackage[english]{babel}
\usepackage{amssymb,amsmath,amsfonts}
\usepackage{graphicx}
\usepackage{hyperref}

\usepackage[usenames,dvipsnames]{color}

\usepackage[mathlines]{lineno}

\def\bra#1{\langle #1|}
\def\ket#1{|\mbox{$#1$}\rangle}

\def\Z{
\begin{bmatrix}
  z_{\kappa_m,k_n,\tau} \\
  1
\end{bmatrix}}
\def\Zp{
\begin{bmatrix}
  z_{\kappa_m,-k_n,\tau} \\
  1
\end{bmatrix}}

\begin{document}

\title{\Large Staggered spin-orbit interaction in a nanoscale device}


\author{L.C.~Contamin}
\affiliation{Laboratoire de Physique de l'\'{E}cole Normale Sup\'{e}rieure, ENS, Universit\'{e} PSL, CNRS, Sorbonne Universit\'{e}, Universit\'{e} Paris-Diderot, Sorbonne Paris Cit\'{e}, Paris, France.}

\author{T.~Cubaynes}
\affiliation{Laboratoire de Physique de l'\'{E}cole Normale Sup\'{e}rieure, ENS, Universit\'{e} PSL, CNRS, Sorbonne Universit\'{e}, Universit\'{e} Paris-Diderot, Sorbonne Paris Cit\'{e}, Paris, France.}

\author{W.~Legrand}
\affiliation{Laboratoire de Physique de l'\'{E}cole Normale Sup\'{e}rieure, ENS, Universit\'{e} PSL, CNRS, Sorbonne Universit\'{e}, Universit\'{e} Paris-Diderot, Sorbonne Paris Cit\'{e}, Paris, France.}

\author{M.~Marganska}
\affiliation{Institute for Theoretical Physics, University of Regensburg, 93040 Regensburg, Germany.}

\author{M.M.~Desjardins}
\affiliation{Laboratoire de Physique de l'\'{E}cole Normale Sup\'{e}rieure, ENS, Universit\'{e} PSL, CNRS, Sorbonne Universit\'{e}, Universit\'{e} Paris-Diderot, Sorbonne Paris Cit\'{e}, Paris, France.}

\author{M.~Dartiailh}
\affiliation{Laboratoire de Physique de l'\'{E}cole Normale Sup\'{e}rieure, ENS, Universit\'{e} PSL, CNRS, Sorbonne Universit\'{e}, Universit\'{e} Paris-Diderot, Sorbonne Paris Cit\'{e}, Paris, France.}

\author{Z.~Leghtas}
\affiliation{Laboratoire de Physique de l'\'{E}cole Normale Sup\'{e}rieure, ENS, Universit\'{e} PSL, CNRS, Sorbonne Universit\'{e}, Universit\'{e} Paris-Diderot, Sorbonne Paris Cit\'{e}, Paris, France.}
\affiliation{QUANTIC team, INRIA de Paris, Paris, France.}
\affiliation{Centre Automatique et Systèmes, Mines-ParisTech, PSL Research University, Paris, France.}

\author{A.~Thiaville}
\affiliation{Laboratoire de Physique des Solides, Université Paris-Saclay, CNRS UMR 8502, Orsay, France.}

\author{S.~Rohart}
\affiliation{Laboratoire de Physique des Solides, Universit'{e} Paris-Saclay, CNRS UMR 8502, Orsay, France.}

\author{A.~Cottet}
\affiliation{Laboratoire de Physique de l'\'{E}cole Normale Sup\'{e}rieure, ENS, Universit\'{e} PSL, CNRS, Sorbonne Universit\'{e}, Universit\'{e} Paris-Diderot, Sorbonne Paris Cit\'{e}, Paris, France.}

\author{M.R.~Delbecq}
\email[email: ]{matthieu.delbecq@ens.fr}
\affiliation{Laboratoire de Physique de l'\'{E}cole Normale Sup\'{e}rieure, ENS, Universit\'{e} PSL, CNRS, Sorbonne Universit\'{e}, Universit\'{e} Paris-Diderot, Sorbonne Paris Cit\'{e}, Paris, France.}
\thanks{These authors co-supervised this work.}

\author{T.~Kontos}
\email[email: ]{takis.kontos@ens.fr}
\affiliation{Laboratoire de Physique de l'\'{E}cole Normale Sup\'{e}rieure, ENS, Universit\'{e} PSL, CNRS, Sorbonne Universit\'{e}, Universit\'{e} Paris-Diderot, Sorbonne Paris Cit\'{e}, Paris, France.}
\thanks{These authors co-supervised this work.}

\date{\today}

\begin{abstract}
The coupling of the spin and the motion of charge carriers stems directly from the atomic structure of a conductor. It has become an important ingredient for the emergence of topological matter, and, in particular, topological superconductivity which could host non-abelian excitations such as Majorana modes or parafermions. These modes are sought after mostly in semiconducting platforms which are made of heavy atoms and therefore exhibit naturally a large spin-orbit interaction. Creating domain walls in the spin orbit interaction at the nanoscale may turn out to be a crucial resource for engineering topological excitations suitable for universal topological quantum computing. For example, it has been proposed for exploring exotic electronic states \cite{Klinovaja2020a} or for creating hinge states \cite{Klinovaja2020b}. Realizing this in natural platforms remains a challenge. In this work, we show how this can be alternatively implemented by using a synthetic spin orbit interaction induced by two lithographically patterned magnetically textured gates. By using a double quantum dot in a light material- a carbon nanotube- embedded in a microwave cavity, we trigger hopping between two adjacent orbitals with the microwave photons and directly compare the wave functions separated by the domain wall via the light-matter coupling. We show that we can achieve an engineered staggered spin-orbit interaction with a change of strength larger than the hopping energy between the two sites.
\end{abstract}

\maketitle

Recently there has been a growing interest in controlling a crucial handle on the spin, the spin orbit interaction. It is a particularly important resource for spin qubits \cite{ Viennot2015, Mi2018a, Samkharaze2018,Nadj-Perge2010} but also for the engineering of topological states \cite{Oreg2010, Lutchyn2010}. Most of the experiments carried out so far rely on transport measurements, which do not enable a direct probe of the spatial structure of the spin-orbit interaction. Whereas local probes such as STM \cite{Nadj-Perge2013, Jack2019} are a priori very well suited to perform such a task, they remain challenging to implement in quantum devices \cite{Lesueur2008}. An ideal setup for spatial resolution is two localized orbitals which experience two different spin orbit interactions separated by a spin-independent tunnel barrier. The overlap of the two corresponding wave functions can be directly mapped onto the tunnel matrix element through the barrier. Such a setup can readily be implemented using a double quantum dot and the corresponding matrix element can directly be measured by cavity quantum electrodynamics techniques \cite{Viennot2015}.

It might seem counterintuitive at first glance that microwaves which have macroscopic wavelengths can probe such nanoscale features. We show here that the large electric field gradients which can be achieved inside a microwave cavity enable nanoscale dipoles to be sensed \cite{Desjardins2017}. Specifically, we study a device made out of a carbon nanotube double quantum dot proximal to two different magnetic textures inducing locally different synthetic spin orbit interactions \cite{Desjardins2019}. As a consequence, the localized energy levels respond differently to the external magnetic field. The phase contrast of the microwave signal reveals a large difference of the magnetic field response of the two dots, witnessing a large spin orbit contrast at the nanoscale.

\begin{figure}\centering
\includegraphics[width=0.7\textwidth]{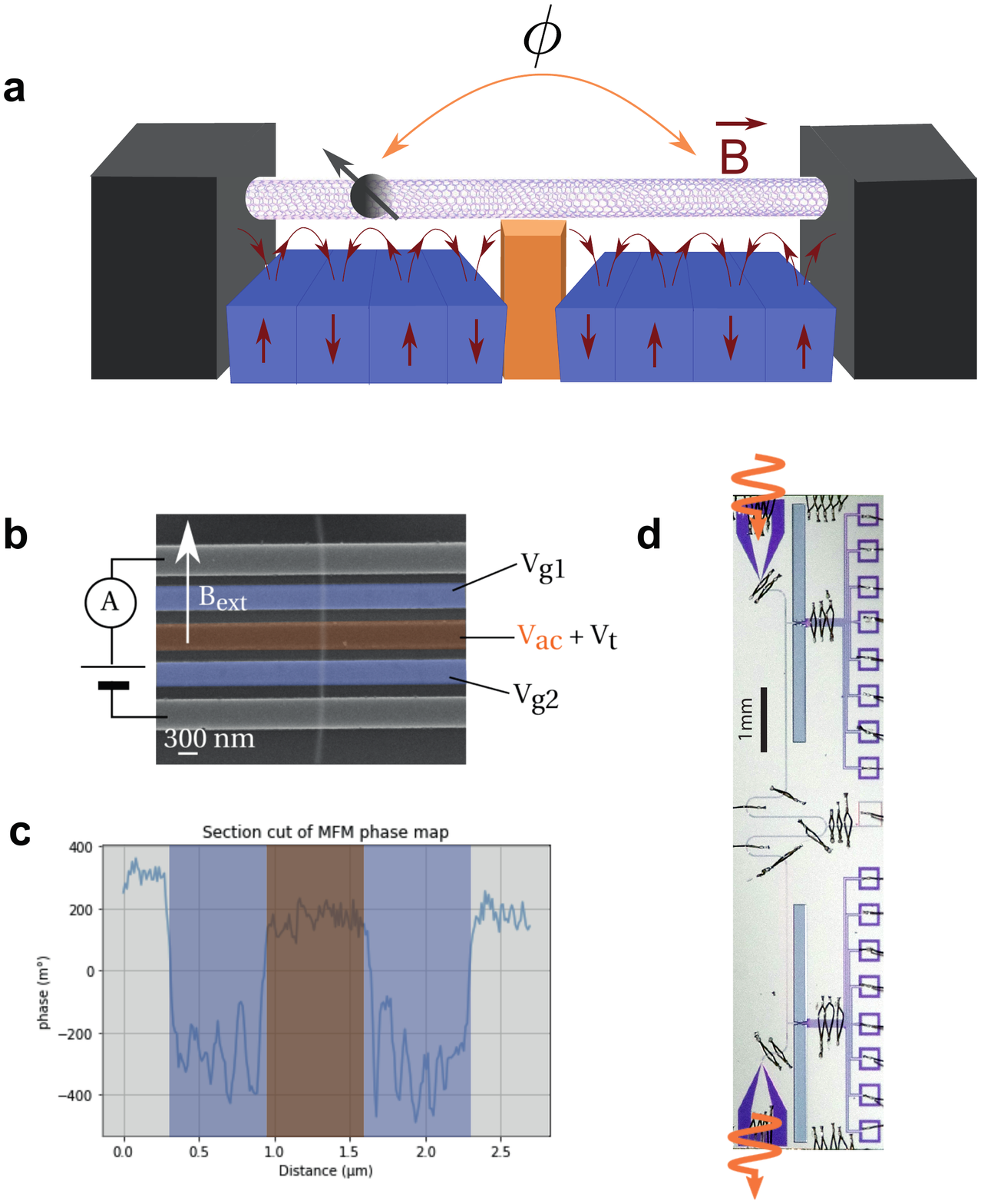}
\caption{\textbf{The device} (a) Schematics of the DQD. It is defined inside a CNT connected to two normal contacts (grey). The two blue gates yield an oscillating magnetic field (red) at the level of the dots. The DQD is also coupled to a microwave cavity through the orange gate; as a consequence, the transmission phase $\Phi$ is sensitive to the charge susceptibility of the DQD. (b) False-color SEM image of sample 2, with the CNT highlighted in white. The central orange gate is connected to the cavity central conductor. The two lateral blue gates are made of teh CoPt stacks. (c) MFM phase cut of the sample, displaying oscillations of the AFM phase signal above the magnetic gates (blue region). (d) Photograph of the CPW in which the sample is embedded.}
\label{fig:setup}
\end{figure}

The principle of our experiment is depicted in figure~\ref{fig:setup}a. A double quantum dot with each of the two dots subject to two different synthetic spin orbit interactions is coupled to a photonic field which actuates tunneling between these two dots as schematized by the orange arrow \cite{Kloeffel2013}. The electric dipole $\phi$ arising from tunneling between the two dots stems directly from the overlap between the wave functions between the left and the right dot depicted here by a specific effective spin direction on the left dot. Owing to the band structure of carbon nanotubes, the Hilbert space has at least 4 dimensions for each of the two dots due to the spin and the orbital degrees of freedom.

Our physical implementation of such a setup is presented in figure~\ref{fig:setup}b. A double quantum dot is patterned in a single wall carbon nanotube using a stapling technique \cite{Cubaynes2019}. The device is designed with two magnetically textured gates, colored in blue in figure~\ref{fig:setup}b, made out of CoPt stacks (see Appendix) \cite{Desjardins2019} above which the nanotube is stapled. The device is embedded in a Nb microwave cavity with a quality factor of about $1000$ and a resonance frequency of $f_{cav} = \unit{6.42}{\giga\hertz}$, shown in figure~\ref{fig:setup}d. From the magnetic force microscope (MFM) micrograph shown in figure~\ref{fig:setup}c, modulations of the magnetic signal are observed with a length scale $\lambda$ of about $200$ nm. This yields a priori a large spin orbit energy scale \cite{Klinovaja2012,Kloeffel2013} of $h v_F /2 \lambda \approx \unit{8}{\milli\electronvolt}$, where $v_F$ is the Fermi velocity in the SWNT, comparable to the mean energy level spacing of each dot , of about $h v_F /2 L \approx \unit{3}{\milli\electronvolt}$, where $L\approx500$ nm is the designed physical length of each dot.

\begin{figure}\centering
\includegraphics[width=0.8\textwidth]{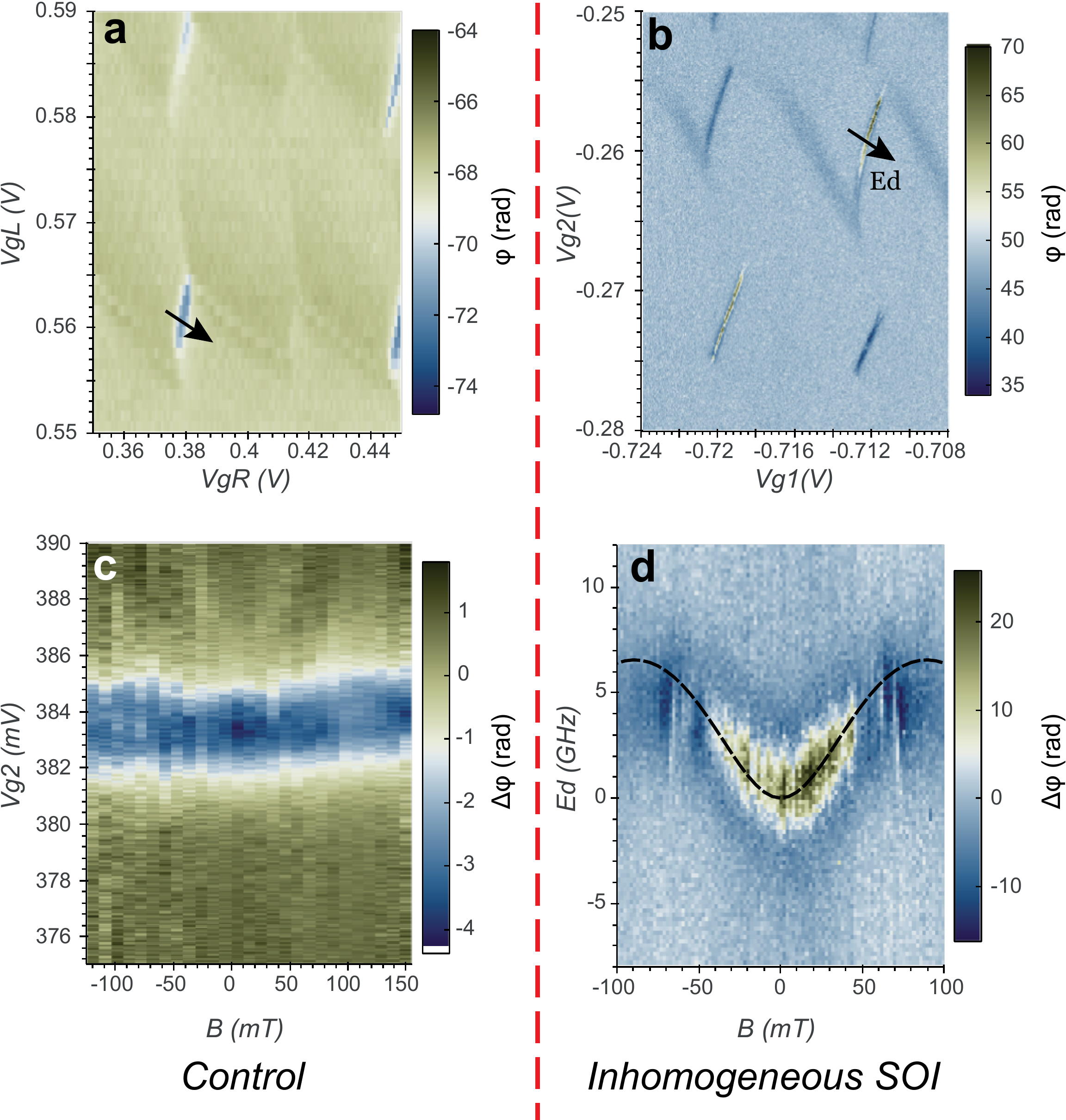}
\caption{\textbf{Magneto-spectroscopy of staggered and control device}.
\textbf{a,b} $V_{g1}-V_{g2}$ cavity phase shift map for two samples. \textbf{c,d} $B-V_g$ maps of the cavity phase shift, over a magnetic field range of about $\unit{\pm 0.1}{\tesla}$ for the two samples. For sample 1 \textbf{c} the cavity signal is unaffected by the external magnetic field, as opposed to sample 2 \textbf{d}. For these measurements, the cavity resonant frequency is measured at each new value of external magnetic field.}
\label{fig:bat}
\end{figure}

\begin{figure}\centering
\includegraphics[width=0.7\textwidth]{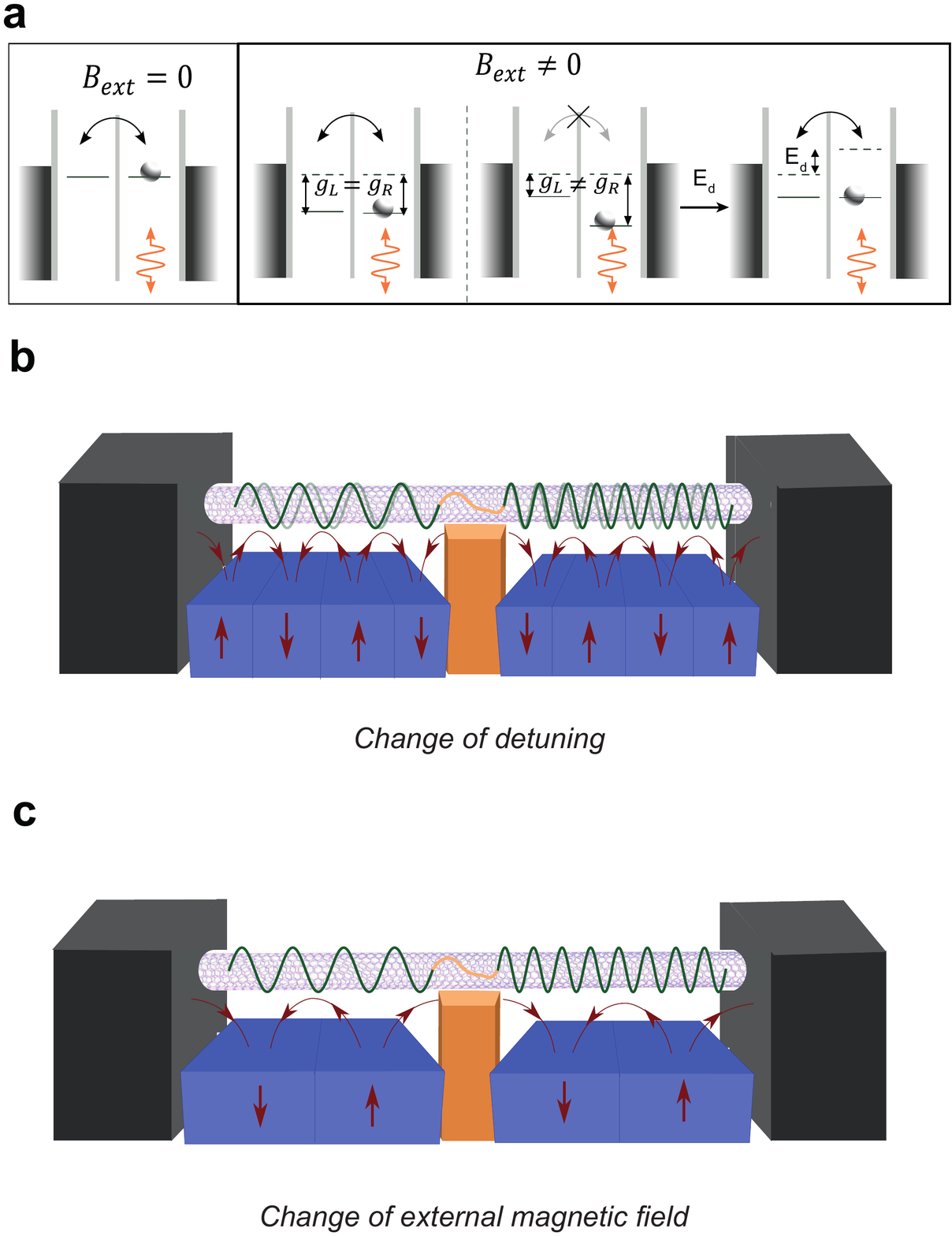}
\caption{\textbf{Picture of evolution of DQD levels and of the staggered spin-orbit interaction}. \textbf{a} Schematic representation of the evolution of the charge susceptibility with an external field $B_{ext}$, from a no-detuning situation (first panel from the left). When the left and right Landé factors are equal, $g_L=g_R$, the cavity signal is unaffected by $B_{ext}$ (second panel). When they are different, the cavity signal is modified (third panel), but the resonant condition can be recovered with a detuning $E_{d}$ (fourth panel). \textbf{b} Picture illustrating the evolution of the spinful levels from the overlap between the electronic wavefunction and the magnetic textures from change of detuning. \textbf{c} Picture illustrating the evolution of the spinful levels from the overlap between the electronic wavefunction and the magnetic textures from change of detuning.}
\label{fig:cartoon}
\end{figure}

 The phase of the cavity transmission is sensitive to the charge susceptibility of the quantum circuit \cite{Cottet2017}; in a double-quantum dot setup, a phase shift is measured when two energy levels are essentially resonant as sketched in the leftmost panel of figure~\ref{fig:cartoon}a. When an external magnetic field is applied, $B_{ext}\neq 0$, the energy levels are shifted by an energy $\delta E_{L(R)} = g_{L(R)} \mu_{B}B_{ext}$ where $g_{L(R)}$ is an effective Land\'e factor in the left (right) dot. In absence of a magnetic texture, and the two dots being formed inside the same CNT, the Land\'e factor is expected to be identical in the two dots and $\delta E_{L}=\delta E_{R}$ as shown in the left panel of figure~\ref{fig:cartoon}a. In this situation, the cavity signal at the resonant frequency is expected to be unaffected by the external magnetic field (the line width of the cavity is not affected much at these magnetic fields). Such a behavior, observed in several samples, is shown in figure~\ref{fig:bat}c. Such a \textit{control} sample design is similar to the one presented in figure~\ref{fig:setup} but its magnetic electrode generates a very small and \textit{non-modulated} dipolar field. Panel a presents the phase contrast as a function of $V_{gR}-V_{gL}$ for the control sample, where both the stability diagram and internal transitions between the two dots are visible. Panel c presents the evolution of the phase contrast as a function of $V_{gR}-B_{ext}$ over the values spanned by the black arrow in panel a. The cavity signal at resonance does not change with an external magnetic field between $\unit{\pm 150}{\milli \tesla}$, as expected for $g_{L}=g_{R}$. A very different signal is observed for the staggered sample, fabricated with different magnetically textured gates. Here, the measurements correspond to a second situation where $g_{L} \neq g_{R}$, as illustrated in the third panel of figure~\ref{fig:cartoon}a. The resonance condition for the cavity signal change with $B_{ext}$, and resonance can be recovered by changing the detuning $E_{d}$, as illustrated in figure~\ref{fig:bat}d. Same as for the control sample, panel b presents the stability diagram of the sample and panel d the evolution of the phase signal with an external magnetic field an a detuning $E_{d}$ (its span being represented by the black arrow in b). Here, the cavity signal changes both in detuning value and in contrast over a range of $B_{ext}=\unit{\pm 100}{\milli \tesla}$. Two qualitative observations can be made here. First, the width of the phase signal gives an order of magnitude of the energy scale at play: it is of the order of $t \approx \hbar \omega_{cav}$ (see Appendix). The change in the resonant condition thus corresponds to a detuning shift ($\delta E_{d}$ in figure~\ref{fig:bat}a) larger than $t$. Second, the change in the phase contrast indicates a change in the effective interdot tunneling term $ t\sin(\theta)$ with $B_{ext}$, where $\theta$ is the angle between the spin eigenvalues for the two energy levels at play. Importantly, the observation that the dispersion of the phase signal occurs on an energy scale $t$ shows that the inhomogeneity of the energy scale governing the dispersion is $t$ i.e. the strong inhomogeneity regime according to the conventional wisdom for a chain.

\begin{figure}\centering
\includegraphics[width=0.8\textwidth]{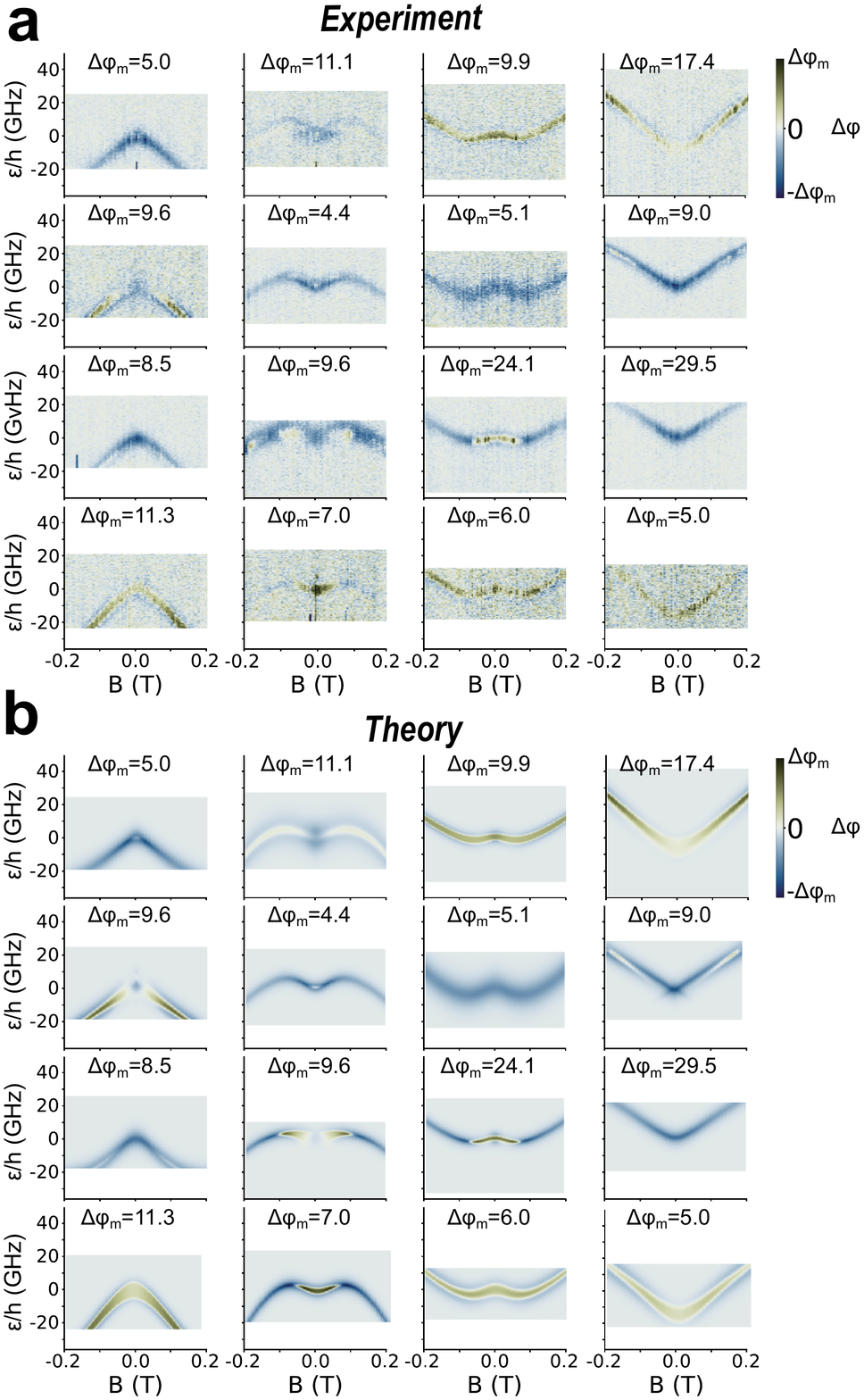}
\caption{\textbf{Orbital effect on spinful transitions}. \textbf{a} Measured $B-V_g$ phase shift maps for several transitions in sample 2. The magnetic field is swept from $\unit{-0.2}{\tesla}$ to $\unit{0.2}{\tesla}$. \textbf{b} Modeling of the transitions with the local fields and effective $g$ model of the Appendix section.}
\label{fig:batlarge}
\end{figure}

In order to further susbtantiate our findings, we study now in details the evolution of the phase contrast dispersion for different orbital states i.e. for different $V_{gR}-V_{gL}$ gate configurations. Figure~\ref{fig:batlarge}a displays the corresponding measurements of the phase contrast as a function of $B_{ext}-E_{d}$ for several transitions in the staggered sample. As expected for an orbitally sensitive phenomenon, there are strong qualitative variations for the phase contrast dispersion depending on the charge states considered. The observed dispersions range from a "v-shape" going up or down to a "w-shape" going up or down, the capsized "w-shape" of figure \ref{fig:bat}d being one particular example. In addition, there are changes in the magnitude and sign of the phase contrasts as a function of the external magnetic field. As a consequence, these measurements show that there are spinful levels the dispersion of which change as the orbital part of the wave function is changed, the hallmark of spin-orbit interaction. Since, alike figure \ref{fig:bat}d, the dispersion is comparable to the width of the phase contrasts stripe, we are led to conclude that the spin-orbit interaction engineered in our setup is in the strongly inhomogeneous regime i.e. has a staggered character between the two dots. This is the main result of our work.

We now show that we can understand quantitatively our findings. In a double dot setup with a large orbital level spacing, our system can be described with an effective spin qubit model \cite{Cottet2010,Kloeffel2013} (see Appendix). The left (right) dot are subject to a local field $B_{L(R)}$ with a relative angle $\theta$ in the x-y plane and has an effective Land\'e factor $g_{L(R)}$. As shown in figure~3b and c, these effective parameters stem from the overlap of the local magnetic field modulations and the electronic wave function. We consider them as a fitting parameter (see Appendix). The external magnetic field $B_{ext}$ is applied along $z$ the CNT axis and $B_{L}$ is taken along the $x$ axis. As shown in the Appendix, all the ingredients of such a low energy model can be obtained by considering the effect of the magnetic texture on a carbon nanotube. In particular, the renormalization of the spinful energy levels acquire a magnetic field dependent part which strongly renormalize the effective Landé factors. The magnetic field dispersion of the spinful levels in each dot being related to the overlap between the electronic wave function and magnetic texture, there are two ways to control them, either by detuning each dot or by changing the external magnetic field,  as illustrated in figure~3b and c. As shown in figure~\ref{fig:batlarge}b, we can quantitatively reproduce the variety of experimental observations. The obtained orders of magnitude of $B_{L(R)}$, $\theta$ and $g_{L(R)}$ are $100mT-500 mT$, $0-0.95\pi $, $1-200$ and $t\approx 3-7 GHz$ (see Appendix). Besides the qualitative dependence on the orbital and inhomogeneity, it is worth noticing that the extracted values of Landé factors are much larger than the observed values in the litterature and in our control devices. This allows us to rule out a simple orbital effect for our measurements. All these facts confirm that we have achieved the strong inhomogeneity regime of spin orbit interaction.

As a conclusion, we have demonstrated that, by using a magnetic texture, we can achieve changes in the magnitude and direction of the spin orbit interaction which correspond to an energy larger than the hopping between adjacent orbitals. Such a synthetic material could have important applications for the engineering of topologically non-trivial states as well as for designs of spin quantum bits.


\textbf{APPENDIX}

\textbf{Fabrication and measurement techniques.}

The two sample presented in the main text consist in DQD made out of a CNT, stapled over a mesoscopic
circuit using the stapling technique described in ref \cite{Cubaynes2019}. The DQD is coupled to a CPW resonator etched from a Nb thin film.

We describe here in greater details the fabrication and measurement techniques. The electrical circuit and microwave cavity were lithographically defined on a high-resistivity Si/Si02 substrate. The cavity is made of a $\lambda/2$ coplanar waveguide (CPW). First, a 100nm-thick Nb layer is evaporated at a pressure below $\unit{5\times 10^{-10}}{\milli\bbar}$, then the cavity pattern is defined with laser lithography and etched using a reactive ion etching (RIE) process with $SF_6$. The CPW of the staggered sample is represented in figure~\ref{fig:setup} d, and has a resonant frequency around $\unit{6.42}{\giga\hertz}$. Then, the nanoscale circuit for defining the DQD is drawn using electron beam lithography and metal evaporation processes. Trenches are defined around this circuit with either optical or electron beam lithography and RIE etching.
Carbon nanotubes were chemically grown using a methane process, on a separated chip designed for the stapling process, and subsequently stapled under vacuum. Once an good electrical contact is measured at room temperature, the circuit is transferred to a cryostat.

Both samples were characterized in a dilution fridge with base temperature of about $\unit{20}{\milli\kelvin}$, through simultaneous DC and RF measurements. For RF measurement, a heterodyne detection scheme is used with a modulation frequency of $\unit{20}{\mega\hertz}$. For every change in magnetic field, the change in the bare cavity resonant frequency is measured with the DQD transitions detuned.


The two samples differ in the nanoscale circuit defining the DQD. For the staggered sample, the CNT was positioned over two magnetically textured gates and a central Al/Alox gate. This last gate can be DC biased and is also connected to the central conductor of the microwave cavity of resonant frequency $f_{cav} = \unit{6.42}{\giga\hertz}$ and linewidth $\kappa = \unit{5.5}{\mega\hertz}$. The magnetic gates were made out of ten repetitions of Co/Pt, with a Ta/Pt initial layer and a thin Alox cap. The CNT was connected to two Pd electrodes, through which a current can be measured. The electrode height is chosen so that the CNT was lying on the magnetic gates, to maximize their effect.
The control sample on the other hand, was fabricated with several Al/Alox gates and two narrow magnetic gates with only 5 repetitions of a Pt/Co bi-layer. The CNT was again connected with two Pd contacts. The electrode height was increased, to that the CNT was suspended above the gate structure (as in ref \cite{Cubaynes2019}). The CNT was capacitively coupled to a microwave cavity through one of the Al/Alox gate, of resonant frequency $\unit{6.439}{\giga\hertz}$ and quality factor $1600$. The narrower magnetic gates in this control case implied that we had essentially a single domain situation (or bi-domain at most) ensuring that there was no magnetic texture, as shown by magnetic force microscopy measurements. 

\textbf{Low energy hamiltonian of a carbon nanotube in the presence of a magnetic texture}
We present in this section the derivation of the low energy hamiltonian of a single wall carbon nanotube in the presence of a magnetic texture. The spectrum of the SWNT subject to an external magnetic field reads \cite{Bulaev08}:
\begin{equation}
E_{\kappa,k,\tau,\sigma}=\pm \sqrt{\kappa^2+k^2}+\frac{1}{2}g_{orb}B_{\parallel}\tau+\frac{1}{2}g_{s}B_{ext}\sigma
\end{equation}
where $g_{orb(s)}$ are the orbital (spin) Land\'e factors, $\tau(\sigma)$ are the orbital(spin) indices, $\kappa$ and $k$ are the transverse and longitudinal wave vectors of the nanotube. Using the conventional quantization conditions for both $\kappa$ and $k$, we can introduce the wave functions of electrons/holes in a quantum dot made out of a carbon nanotube:
\begin{equation}
\langle\varphi,\zeta|\Psi_{m,n,\tau,\sigma}\rangle=\frac{e^{i\overrightarrow{K}(').\overrightarrow{r}}}{\sqrt{4\pi}}e^{i(m-\tau\nu/3)\varphi}\Phi_{m,n}(\zeta)
\end{equation}
where $n,m$ are the quantum numbers for the transverse and longitudinal quantization. The parameter $\nu=0,\pm 1$ encodes whether the nantoube is semiconducting ($\nu = \pm 1$) or metallic ($\nu = 0$). The wave function $\Phi_{m,n}(\zeta)$ has the usual spinor structure to account for the graphene sublattices \cite{Bulaev08}:
\begin{equation}
\Phi_{m,n}(\zeta)=C \Z e^{ik_n \zeta}+D \Zp e^{-ik_n \zeta}
\end{equation}
with $z_{\kappa,k,\tau}=\pm \tau(\kappa-i\tau k)/\sqrt{\kappa^2+k^2}$. The coefficients $C$ and $D$ depend on the boundary conditions of the nanotube.
We would like to calculate the matrix elements arising from the spin texture. The corresponding terms in the nanotube hamiltonian read\cite{Egger12}:
\begin{align*}
  spin &: \frac{1}{2}g_{s}B_{osc}(\hat{\sigma}_z \cos 2\pi\zeta/\lambda+\hat{\sigma}_x \sin 2\pi\zeta/\lambda)\\
  valley &: \frac{1}{2}g_{orb}B_{osc} \hat{\tau}_z \hat{\eta}_x \cos 2\pi\zeta/\lambda
\end{align*}
In the above expressions, we have assumed a cycloidal magnetic texture oscillating with a period $\lambda$ and an amplitude $B_{osc}$. The matrices $\hat{\sigma}_i$, $\hat{\tau}_i$ and $\hat{\eta}_i$ are the Pauli matrices acting on the spin, valley and sublattice spaces respectively. We define $k_{\lambda}=2 \pi/ \lambda$. The matrix element of these terms for the wave functions $|\Psi_{m,n,\tau,\sigma}\rangle$ are all of the form :
\begin{align*}
  \mathcal{A}_{mnn'} \frac{\sin((k_n\pm k_{n'}\pm k_{\lambda})L/2)}{(k_n\pm k_{n'}\pm k_{\lambda})L/2}
\end{align*}
where $L$ is the length of the confined region of the nanotube forming the quantum dot (assuming a square potential for the sake of simplicity) and $\mathcal{A}_{mnn'}$ is a coefficient which depends on the overlap between the wave functions of the dot and the subband index.
We would like now to obtain an effective spin-valley hamiltonian for the CNT. Two terms arise from the above discussion : terms which conserve the longitudinal index (first order) and terms which couple different orbitals.The hamiltonian of the system is now :
\begin{align}\label{effective}
  H=\sum_n |n\rangle\langle n|[E_n+ \frac{1}{2}g_{s}\mu_B (B_{ext}+\alpha^{\sigma_z}_{nn}B_{osc}\hat{\sigma}_z+\alpha^{\sigma_x}_{nn}B_{osc}\hat{\sigma}_x)+\frac{1}{2}g_{orb}\mu_B (B_{ext}+\beta^{\tau_z}_{nn}B_{osc})\hat{\tau}_z]\\+\sum_{nn'} |n\rangle\langle n'|[ \frac{1}{2}g_{s}\mu_B B_{osc}(\alpha^{\sigma_z}_{nn'}\hat{\sigma}_z+\alpha^{\sigma_x}_{nn'}\hat{\sigma}_x)+\frac{1}{2}g_{orb}\mu_B B_{osc}\beta^{\tau_z}_{nn'}\hat{\tau}_z]+h.c.
\end{align}
The second terms modify at the second order the hamiltonian. This can be calculated using a Schrieffer-Wolf transformation:
\begin{align*}
\tilde{H}=e^S H e^{-S}\approx H + [S,H]+\frac{1}{2}[S,[S,H]+ ...
\end{align*}
where S is a anti-hermitian operator. The operator S has to be chosen such that its commutator with the diagonal part of the hamiltonian in the orbital subspace (\ref{effective}) is exactly the opposite of the off-diagonal part (5). One can show that an operator satisfying these conditions has the following matrix elements:
\begin{align*}
\langle n,\sigma,\tau|S|m,\sigma',\tau' \rangle=\\\frac{[\sigma \delta_{\sigma\sigma'}\alpha^{\sigma_z}_{nm}+\sigma \delta_{\overline{\sigma}\sigma'}\alpha^{\sigma_x}_{nm}+\tau\beta^{\tau_z}_{nm}]\delta_{\tau\tau'}}{E_n-E_m+\frac{1}{2}g_{s}\mu_B (B_{ext}+B_{osc}\sqrt{{\alpha^{\sigma_z}_{nn}}^2+{\alpha^{\sigma_x}_{nn}}^2})\sigma-\frac{1}{2}g_{s}\mu_B (B_{ext}+B_{osc}\sqrt{{\alpha^{\sigma_z}_{mm}}^2+{\alpha^{\sigma_x}_{mm}}^2})\sigma'}
\end{align*}
where $\sigma$ is the new quantum number along the quantization axis defined by the external field and the first order terms of the magnetic field. The final version of the hamiltonian (projected on the orbital $|n\rangle$) is :
\begin{align}
H_{eff}=E_n+\frac{1}{2}g_{s}\mu_B (B_{ext}+\alpha^{\sigma_z}_{nn}B_{osc}\hat{\sigma}_z+\alpha^{\sigma_x}_{nn}B_{osc}\hat{\sigma}_x)+\frac{1}{2}g_{orb}\mu_B (B_{ext}+\beta^{\tau_z}_{nn}B_{osc})\hat{\tau}_z+\\+
\gamma^{\sigma_z \tau_z}_{nn}\frac{(g_{s} \mu_B B_{osc})^2}{E_{SO}}\hat{\sigma}_z \hat{\tau}_z+\gamma^{\sigma_x \tau_z}_{nn}\frac{(g_{s} \mu_B B_{osc})^2}{E_{SO}}\hat{\sigma}_x \hat{\tau}_z+\gamma^{\sigma_y}_{nn}\frac{(g_{s} \mu_B B_{osc})^2}{E_{SO}}\hat{\sigma}_y
\end{align}
The dimensionless parameters $\alpha^{\sigma_z}_{nn}$, $\alpha^{\sigma_x}_{nn}$, $\beta^{\tau_z}_{nn}$, $\gamma^{\sigma_x \tau_z}_{nn}$, $\gamma^{\sigma_z \tau_z}_{nn}$ and $\gamma^{\sigma_y}_{nn}$ are of the order of 1 and depend on the wave function, the value of $k_\lambda$ and therefore on the external magnetic field $B_{ext}$ as well. It is important to note that we obtain for the two first terms of the second line the same form as that for the intrinsic spin orbit interaction \cite{Laird2013} in carbon nanotubes which shows that the magnetic texture plays indeed the role of an effective spin orbit interaction. Finally, it is essential to note that the synthetic spin-orbit interaction acts already at first order as an effective magnetic field which depends on the wave functions through the parameters $\alpha^{\sigma_z}_{nn}$, $\alpha^{\sigma_x}_{nn}$ and $\beta^{\tau_z}_{nn}$ which imply in particular an orbital dependent effective field direction for the spin defined by the angle $\theta_n=\arctan [\alpha^{\sigma_x}_{nn}/\alpha^{\sigma_z}_{nn}]$.

\textbf{Local fields and effective-$g$ model}

The above hamiltonian can be further simplified if one considers the ground state and the first excited state. This is fully justified in our case since the cavity is energy selective and filters the transition which is the closest to the cavity frequency $\omega_{cav}/2\pi$. We therefore model our devices as a double quantum dot (DQD) with one level in each dot, with an effective spin degree of freedom corresponding to the ground and excited states of hamilotnian (6) and (7).

The two levels are detuned by an energy $E_{d}$. We introduce $\tau_{0,x,y,z}, \sigma_{0,x,y,z}$ as the Pauli matrices for the left/right and effective spin subspaces. Each dot is subject to a local effective field $B_{L,R}$ that is in the x-y plane ($B_L$ is along the x axis, and $B_R$ has an angle $\theta$ to $B_L$). An external magnetic field $B_{ext}$ can be applied along the z-axis, that is the axis of the CNT. $B_{ext}$ can thus both have a Zeeman and orbital contribution to the spectrum. The orthogonality between $B_{ext}$ and the local effective field ensure a symmetric spectrum with respect to $B_{ext}$, as experimentally observed.

The model hamiltonian is the following :
$$ H_{tot} = \frac{E_{d}}{2} \tau_{z} + t \tau_{x}+ H_{spin, L} + H_{spin, R}$$
with
\begin{align*}
H_{spin, L} &= -\frac{g_{L}\mu_{B}B_{L}}{4} (\tau_{0}+\tau_{z})\sigma_{x} - \frac{g_{L}\mu_{B}B_{ext}}{4} (\tau_{0}+\tau_{z}) \sigma_{z}\\
H_{spin, R} &= -\frac{g_{R}\mu_{B}B_{R}}{4} (\tau_{0}-\tau_{z})(\sin(\theta) \sigma_{x} + \cos(\theta)\sigma_{y}) - \frac{g_{R}\mu_{B}B_{ext}}{4} (\tau_{0}-\tau_{z}) \sigma_{z}
\end{align*}

The cavity transmission is given by:
$T = \frac{\kappa/2}{(f_{cav}-f_{d}) - i\kappa/2 - \chi} $
where $f_{cav}$, $\kappa$ are the cavity resonance frequency and line width and $f_{d}$ is the drive frequency. The charge susceptibility $\chi$ is given by:
$ \chi = \sum_{i,j} \chi_{ij}(n_{i}-n_{j}) \text{ and } \chi_{ij}=\frac{g_{ij}^{2}}{f_{ij}-f_{d} - i(\Gamma_{1} + \Gamma_{\phi}/2)/2}$
with $n_{i}$, $n_{j}$ the thermal occupations at an electronic temperature $T_{e}$, and $f_{ij}=f_{i}-f_{j}$ the transition frequency between eigenvalues i and j. The electron-photon coupling strength $g_{ij}$ is calculated from the electron-photon coupling operator: $g_{ij} = g_{0} | \bra{i} \frac{\tau_{0}-\tau_{z}}{2} \ket{j} | $ where $g_{0}$ is a fitting parameter.
For simplicity, the dephasing rate $\Gamma_{1}=\unit{1}{\mega\hertz}$ is kept constant, whereas the $\Gamma_{\phi}$ for each $(i,j)$ are calculated from the model Hamiltonian $H_{tot}$(defined above) through the projection of the different dephasing operators with the initial and final states. The dephasing operators are $\tau_{z}$ (charge), $\sigma_{z}$ (spin) and $\gamma_{z}$ (valley).

The electronic temperature is taken to be $T_{e}=\unit{150}{\milli\kelvin}$. $g_{orb}=10, \Delta_{KK'}=\unit{6}{\giga\hertz}, \kappa=\unit{5.5}{\mega\hertz}, f_{cav}=f_{d}=\unit{6.42}{\giga\hertz}$. $\Gamma_{\phi, v}=\Gamma_{\phi, s}=\unit{1}{\mega\hertz}$ by default.


An additional model parameter is the conversion energy for the detuning axis. Indeed, a first approximative value for the lever arms of both gates was extracted from Coulomb diamonds. However, the strong interdot coupling deforms the stability diagram and hinders a precise measurement of these lever arms. A correction to the lever arms is thus kept as a general parameter in the model. First, the gate voltages are converted into a detuning value through
$$ \epsilon_{d} = \mu_{1} - \mu_{2} \text{ with } \mu_{1} = 0.31 V_{g1} - 0.025 V_{g2}, \mu_{2} = -0.10 V_{g1} + 0.11 V_{g2} $$
Then, the detuning in the fits is taken as $E_{d}=\epsilon \epsilon_{d}$, where $\epsilon$ is a fit parameter.

\begin{figure}\centering
\includegraphics[width=0.7\textwidth]{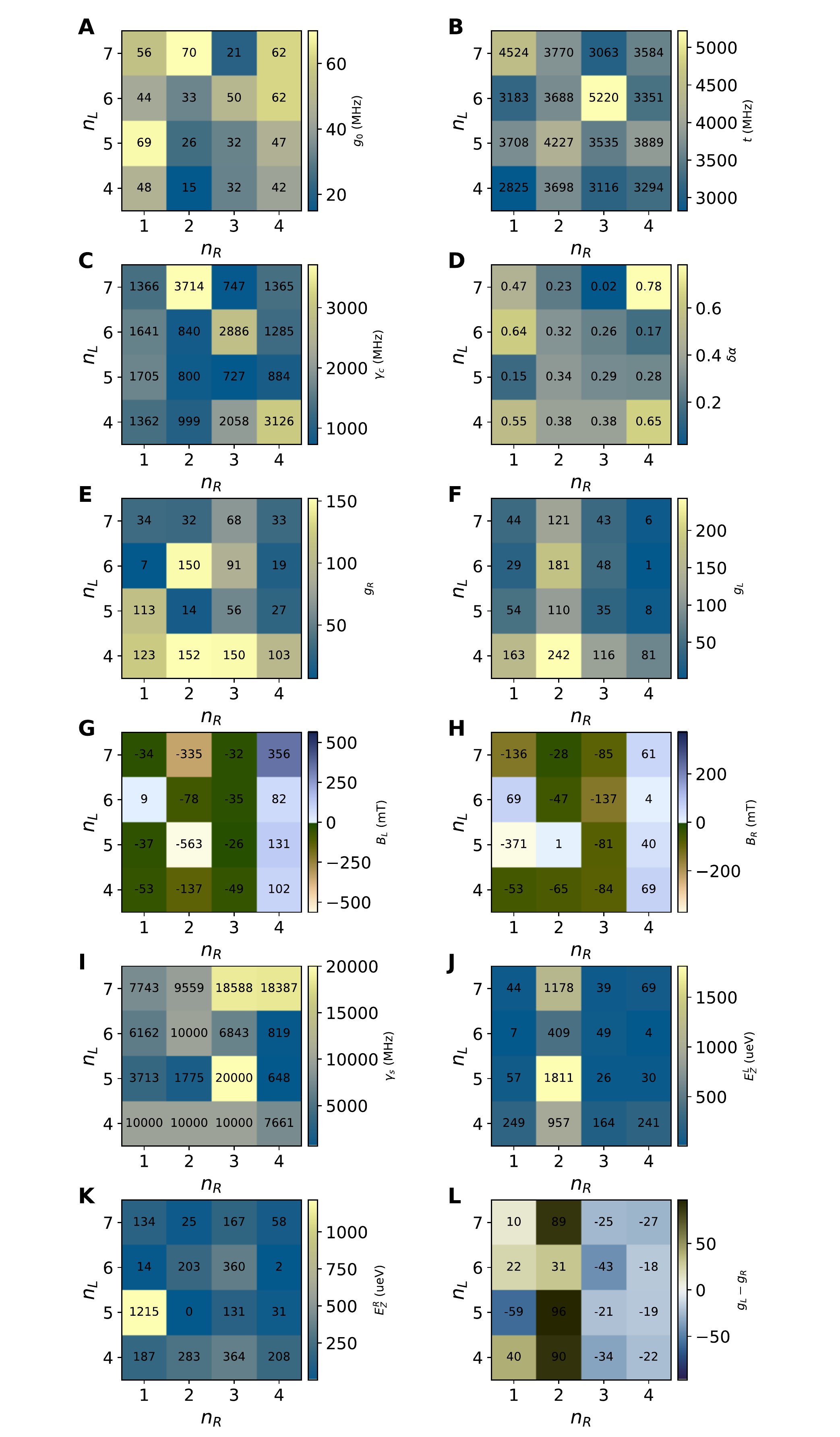}
\caption{\textbf{Fitting parameters}. Summary of all the fitting parameters used in figure~4b}
\label{fig:params}
\end{figure}

The values for the fit of the different transitions are given in figure~\ref{fig:params}.




%

\newpage

\newpage

\newpage

\newpage

\newpage

\newpage






\begin{thebibliography}{21}%
\makeatletter
\providecommand \@ifxundefined [1]{%
 \@ifx{#1\undefined}
}%
\providecommand \@ifnum [1]{%
 \ifnum #1\expandafter \@firstoftwo
 \else \expandafter \@secondoftwo
 \fi
}%
\providecommand \@ifx [1]{%
 \ifx #1\expandafter \@firstoftwo
 \else \expandafter \@secondoftwo
 \fi
}%
\providecommand \natexlab [1]{#1}%
\providecommand \enquote  [1]{``#1''}%
\providecommand \bibnamefont  [1]{#1}%
\providecommand \bibfnamefont [1]{#1}%
\providecommand \citenamefont [1]{#1}%
\providecommand \href@noop [0]{\@secondoftwo}%
\providecommand \href [0]{\begingroup \@sanitize@url \@href}%
\providecommand \@href[1]{\@@startlink{#1}\@@href}%
\providecommand \@@href[1]{\endgroup#1\@@endlink}%
\providecommand \@sanitize@url [0]{\catcode `\\12\catcode `\$12\catcode
  `\&12\catcode `\#12\catcode `\^12\catcode `\_12\catcode `\%12\relax}%
\providecommand \@@startlink[1]{}%
\providecommand \@@endlink[0]{}%
\providecommand \url  [0]{\begingroup\@sanitize@url \@url }%
\providecommand \@url [1]{\endgroup\@href {#1}{\urlprefix }}%
\providecommand \urlprefix  [0]{URL }%
\providecommand \Eprint [0]{\href }%
\providecommand \doibase [0]{http://dx.doi.org/}%
\providecommand \selectlanguage [0]{\@gobble}%
\providecommand \bibinfo  [0]{\@secondoftwo}%
\providecommand \bibfield  [0]{\@secondoftwo}%
\providecommand \translation [1]{[#1]}%
\providecommand \BibitemOpen [0]{}%
\providecommand \bibitemStop [0]{}%
\providecommand \bibitemNoStop [0]{.\EOS\space}%
\providecommand \EOS [0]{\spacefactor3000\relax}%
\providecommand \BibitemShut  [1]{\csname bibitem#1\endcsname}%
\let\auto@bib@innerbib\@empty
\bibitem [{\citenamefont {Ronetti}\ \emph {et~al.}(2020)\citenamefont
  {Ronetti}, \citenamefont {Plekhanov}, \citenamefont {Loss},\ and\
  \citenamefont {Klinovaja}}]{Klinovaja2020a}%
  \BibitemOpen
  \bibfield  {author} {\bibinfo {author} {\bibfnamefont {F.}~\bibnamefont
  {Ronetti}}, \bibinfo {author} {\bibfnamefont {K.}~\bibnamefont {Plekhanov}},
  \bibinfo {author} {\bibfnamefont {D.}~\bibnamefont {Loss}}, \ and\ \bibinfo
  {author} {\bibfnamefont {J.}~\bibnamefont {Klinovaja}},\ }\href@noop {}
  {\bibfield  {journal} {\bibinfo  {journal} {Physical Review Research}\
  }\textbf {\bibinfo {volume} {2}},\ \bibinfo {pages} {022052} (\bibinfo {year}
  {2020})}\BibitemShut {NoStop}%
\bibitem [{\citenamefont {Plekhanov}\ \emph {et~al.}(2020)\citenamefont
  {Plekhanov}, \citenamefont {Ronetti},\ and\ \citenamefont
  {Klinovaja}}]{Klinovaja2020b}%
  \BibitemOpen
  \bibfield  {author} {\bibinfo {author} {\bibfnamefont {K.}~\bibnamefont
  {Plekhanov}}, \bibinfo {author} {\bibfnamefont {D.}~\bibnamefont {Ronetti},
  \bibfnamefont {F.and~Loss}}, \ and\ \bibinfo {author} {\bibfnamefont
  {J.}~\bibnamefont {Klinovaja}},\ }\href@noop {} {\bibfield  {journal}
  {\bibinfo  {journal} {Physical Review Research}\ }\textbf {\bibinfo {volume}
  {2}},\ \bibinfo {pages} {013083} (\bibinfo {year} {2020})}\BibitemShut
  {NoStop}%
\bibitem [{\citenamefont {Viennot}\ \emph {et~al.}(2015)\citenamefont
  {Viennot}, \citenamefont {Dartiailh}, \citenamefont {Cottet},\ and\
  \citenamefont {Kontos}}]{Viennot2015}%
  \BibitemOpen
  \bibfield  {author} {\bibinfo {author} {\bibfnamefont {J.~J.}\ \bibnamefont
  {Viennot}}, \bibinfo {author} {\bibfnamefont {M.~C.}\ \bibnamefont
  {Dartiailh}}, \bibinfo {author} {\bibfnamefont {A.}~\bibnamefont {Cottet}}, \
  and\ \bibinfo {author} {\bibfnamefont {T.}~\bibnamefont {Kontos}},\ }\href
  {\doibase 10.1126/science.aaa3786} {\bibfield  {journal} {\bibinfo  {journal}
  {Science}\ }\textbf {\bibinfo {volume} {349}},\ \bibinfo {pages} {408}
  (\bibinfo {year} {2015})}\BibitemShut {NoStop}%
\bibitem [{\citenamefont {Mi}\ \emph {et~al.}(2018)\citenamefont {Mi},
  \citenamefont {Benito}, \citenamefont {Putz}, \citenamefont {Zajac},
  \citenamefont {Taylor}, \citenamefont {Burkard},\ and\ \citenamefont
  {Petta}}]{Mi2018a}%
  \BibitemOpen
  \bibfield  {author} {\bibinfo {author} {\bibfnamefont {X.}~\bibnamefont
  {Mi}}, \bibinfo {author} {\bibfnamefont {M.}~\bibnamefont {Benito}}, \bibinfo
  {author} {\bibfnamefont {S.}~\bibnamefont {Putz}}, \bibinfo {author}
  {\bibfnamefont {D.~M.}\ \bibnamefont {Zajac}}, \bibinfo {author}
  {\bibfnamefont {J.~M.}\ \bibnamefont {Taylor}}, \bibinfo {author}
  {\bibfnamefont {G.}~\bibnamefont {Burkard}}, \ and\ \bibinfo {author}
  {\bibfnamefont {J.~R.}\ \bibnamefont {Petta}},\ }\href {\doibase
  10.1038/nature25769} {\bibfield  {journal} {\bibinfo  {journal} {Nature}\
  }\textbf {\bibinfo {volume} {555}},\ \bibinfo {pages} {599} (\bibinfo {year}
  {2018})}\BibitemShut {NoStop}%
\bibitem [{\citenamefont {Samkharaze}\ and\ \citenamefont
  {Vandersypen}(2018)}]{Samkharaze2018}%
  \BibitemOpen
  \bibfield  {author} {\bibinfo {author} {\bibfnamefont {S.}~\bibnamefont
  {Samkharaze}}\ and\ \bibinfo {author} {\bibfnamefont {L.~M.~K.}\ \bibnamefont
  {Vandersypen}},\ }\href {\doibase 10.1103/RevModPhys.79.1217} {\bibfield
  {journal} {\bibinfo  {journal} {Science}\ }\textbf {\bibinfo {volume} {79}},\
  \bibinfo {pages} {1217} (\bibinfo {year} {2018})}\BibitemShut {NoStop}%
\bibitem [{\citenamefont {Nadj-Perge}\ \emph {et~al.}(2010)\citenamefont
  {Nadj-Perge}, \citenamefont {Frolov}, \citenamefont {Bakkers},\ and\
  \citenamefont {Kouwenhoven}}]{Nadj-Perge2010}%
  \BibitemOpen
  \bibfield  {author} {\bibinfo {author} {\bibfnamefont {S.}~\bibnamefont
  {Nadj-Perge}}, \bibinfo {author} {\bibfnamefont {S.~M.}\ \bibnamefont
  {Frolov}}, \bibinfo {author} {\bibfnamefont {E.~P. A.~M.}\ \bibnamefont
  {Bakkers}}, \ and\ \bibinfo {author} {\bibfnamefont {L.~P.}\ \bibnamefont
  {Kouwenhoven}},\ }\href {\doibase 10.1038/nature09682} {\bibfield  {journal}
  {\bibinfo  {journal} {Nature}\ }\textbf {\bibinfo {volume} {468}},\ \bibinfo
  {pages} {1084} (\bibinfo {year} {2010})}\BibitemShut {NoStop}%
\bibitem [{\citenamefont {Oreg}\ \emph {et~al.}(2010)\citenamefont {Oreg},
  \citenamefont {Refael},\ and\ \citenamefont {von Oppen}}]{Oreg2010}%
  \BibitemOpen
  \bibfield  {author} {\bibinfo {author} {\bibfnamefont {Y.}~\bibnamefont
  {Oreg}}, \bibinfo {author} {\bibfnamefont {G.}~\bibnamefont {Refael}}, \ and\
  \bibinfo {author} {\bibfnamefont {F.}~\bibnamefont {von Oppen}},\ }\href
  {\doibase 10.1103/PhysRevLett.105.177002} {\bibfield  {journal} {\bibinfo
  {journal} {Physical Review Letters}\ }\textbf {\bibinfo {volume} {105}},\
  \bibinfo {pages} {177002} (\bibinfo {year} {2010})}\BibitemShut {NoStop}%
\bibitem [{\citenamefont {Lutchyn}\ \emph {et~al.}(2010)\citenamefont
  {Lutchyn}, \citenamefont {Sau},\ and\ \citenamefont {{Das
  Sarma}}}]{Lutchyn2010}%
  \BibitemOpen
  \bibfield  {author} {\bibinfo {author} {\bibfnamefont {R.~M.}\ \bibnamefont
  {Lutchyn}}, \bibinfo {author} {\bibfnamefont {J.~D.}\ \bibnamefont {Sau}}, \
  and\ \bibinfo {author} {\bibfnamefont {S.}~\bibnamefont {{Das Sarma}}},\
  }\href {\doibase 10.1103/PhysRevLett.105.077001} {\bibfield  {journal}
  {\bibinfo  {journal} {Physical Review Letters}\ }\textbf {\bibinfo {volume}
  {105}},\ \bibinfo {pages} {077001} (\bibinfo {year} {2010})}\BibitemShut
  {NoStop}%
\bibitem [{\citenamefont {Nadj-Perge}\ \emph {et~al.}(2013)\citenamefont
  {Nadj-Perge}, \citenamefont {Drozdov}, \citenamefont {Bernevig},\ and\
  \citenamefont {Yazdani}}]{Nadj-Perge2013}%
  \BibitemOpen
  \bibfield  {author} {\bibinfo {author} {\bibfnamefont {S.}~\bibnamefont
  {Nadj-Perge}}, \bibinfo {author} {\bibfnamefont {I.~K.}\ \bibnamefont
  {Drozdov}}, \bibinfo {author} {\bibfnamefont {B.~A.}\ \bibnamefont
  {Bernevig}}, \ and\ \bibinfo {author} {\bibfnamefont {A.}~\bibnamefont
  {Yazdani}},\ }\href {\doibase 10.1103/PhysRevB.88.020407} {\bibfield
  {journal} {\bibinfo  {journal} {Physical Review B}\ }\textbf {\bibinfo
  {volume} {88}},\ \bibinfo {pages} {020407} (\bibinfo {year}
  {2013})}\BibitemShut {NoStop}%
\bibitem [{\citenamefont {J{\"{a}}ck}\ \emph {et~al.}(2019)\citenamefont
  {J{\"{a}}ck}, \citenamefont {Xie}, \citenamefont {Li}, \citenamefont {Jeon},
  \citenamefont {Bernevig},\ and\ \citenamefont {Yazdani}}]{Jack2019}%
  \BibitemOpen
  \bibfield  {author} {\bibinfo {author} {\bibfnamefont {B.}~\bibnamefont
  {J{\"{a}}ck}}, \bibinfo {author} {\bibfnamefont {Y.}~\bibnamefont {Xie}},
  \bibinfo {author} {\bibfnamefont {J.}~\bibnamefont {Li}}, \bibinfo {author}
  {\bibfnamefont {S.}~\bibnamefont {Jeon}}, \bibinfo {author} {\bibfnamefont
  {B.~A.}\ \bibnamefont {Bernevig}}, \ and\ \bibinfo {author} {\bibfnamefont
  {A.}~\bibnamefont {Yazdani}},\ }\href {\doibase 10.1126/science.aax1444}
  {\bibfield  {journal} {\bibinfo  {journal} {Science}\ }\textbf {\bibinfo
  {volume} {364}},\ \bibinfo {pages} {1255} (\bibinfo {year}
  {2019})}\BibitemShut {NoStop}%
\bibitem [{\citenamefont {le~Sueur}\ \emph {et~al.}(2008)\citenamefont
  {le~Sueur}, \citenamefont {Joyez}, \citenamefont {Pothier}, \citenamefont
  {Urbina},\ and\ \citenamefont {Esteve}}]{Lesueur2008}%
  \BibitemOpen
  \bibfield  {author} {\bibinfo {author} {\bibfnamefont {H.}~\bibnamefont
  {le~Sueur}}, \bibinfo {author} {\bibfnamefont {P.}~\bibnamefont {Joyez}},
  \bibinfo {author} {\bibfnamefont {H.}~\bibnamefont {Pothier}}, \bibinfo
  {author} {\bibfnamefont {C.}~\bibnamefont {Urbina}}, \ and\ \bibinfo {author}
  {\bibfnamefont {D.}~\bibnamefont {Esteve}},\ }\href@noop {} {\bibfield
  {journal} {\bibinfo  {journal} {Physical Review Letters}\ }\textbf {\bibinfo
  {volume} {100}},\ \bibinfo {pages} {197002} (\bibinfo {year}
  {2008})}\BibitemShut {NoStop}%
\bibitem [{\citenamefont {Desjardins}\ \emph {et~al.}(2017)\citenamefont
  {Desjardins}, \citenamefont {Viennot}, \citenamefont {Dartiailh},
  \citenamefont {Bruhat}, \citenamefont {Delbecq}, \citenamefont {Lee},
  \citenamefont {Choi}, \citenamefont {Cottet},\ and\ \citenamefont
  {Kontos}}]{Desjardins2017}%
  \BibitemOpen
  \bibfield  {author} {\bibinfo {author} {\bibfnamefont {M.~M.}\ \bibnamefont
  {Desjardins}}, \bibinfo {author} {\bibfnamefont {J.~J.}\ \bibnamefont
  {Viennot}}, \bibinfo {author} {\bibfnamefont {M.~C.}\ \bibnamefont
  {Dartiailh}}, \bibinfo {author} {\bibfnamefont {L.~E.}\ \bibnamefont
  {Bruhat}}, \bibinfo {author} {\bibfnamefont {M.~R.}\ \bibnamefont {Delbecq}},
  \bibinfo {author} {\bibfnamefont {M.}~\bibnamefont {Lee}}, \bibinfo {author}
  {\bibfnamefont {M.-S.}\ \bibnamefont {Choi}}, \bibinfo {author}
  {\bibfnamefont {A.}~\bibnamefont {Cottet}}, \ and\ \bibinfo {author}
  {\bibfnamefont {T.}~\bibnamefont {Kontos}},\ }\href {\doibase
  10.1038/nature21704} {\bibfield  {journal} {\bibinfo  {journal} {Nature}\
  }\textbf {\bibinfo {volume} {545}},\ \bibinfo {pages} {71} (\bibinfo {year}
  {2017})}\BibitemShut {NoStop}%
\bibitem [{\citenamefont {Desjardins}\ \emph {et~al.}(2019)\citenamefont
  {Desjardins}, \citenamefont {Contamin}, \citenamefont {Delbecq},
  \citenamefont {Dartiailh}, \citenamefont {Bruhat}, \citenamefont {Cubaynes},
  \citenamefont {Viennot}, \citenamefont {Mallet}, \citenamefont {Rohart},
  \citenamefont {Thiaville}, \citenamefont {Cottet},\ and\ \citenamefont
  {Kontos}}]{Desjardins2019}%
  \BibitemOpen
  \bibfield  {author} {\bibinfo {author} {\bibfnamefont {M.~M.}\ \bibnamefont
  {Desjardins}}, \bibinfo {author} {\bibfnamefont {L.~C.}\ \bibnamefont
  {Contamin}}, \bibinfo {author} {\bibfnamefont {M.~R.}\ \bibnamefont
  {Delbecq}}, \bibinfo {author} {\bibfnamefont {M.~C.}\ \bibnamefont
  {Dartiailh}}, \bibinfo {author} {\bibfnamefont {L.~E.}\ \bibnamefont
  {Bruhat}}, \bibinfo {author} {\bibfnamefont {T.}~\bibnamefont {Cubaynes}},
  \bibinfo {author} {\bibfnamefont {J.~J.}\ \bibnamefont {Viennot}}, \bibinfo
  {author} {\bibfnamefont {F.}~\bibnamefont {Mallet}}, \bibinfo {author}
  {\bibfnamefont {S.}~\bibnamefont {Rohart}}, \bibinfo {author} {\bibfnamefont
  {A.}~\bibnamefont {Thiaville}}, \bibinfo {author} {\bibfnamefont
  {A.}~\bibnamefont {Cottet}}, \ and\ \bibinfo {author} {\bibfnamefont
  {T.}~\bibnamefont {Kontos}},\ }\href {\doibase 10.1038/s41563-019-0457-6}
  {\bibfield  {journal} {\bibinfo  {journal} {Nature Materials}\ }\textbf
  {\bibinfo {volume} {18}},\ \bibinfo {pages} {1060–} (\bibinfo {year}
  {2019})}\BibitemShut {NoStop}%
\bibitem [{\citenamefont {Kloeffel}\ \emph {et~al.}(2013)\citenamefont
  {Kloeffel}, \citenamefont {Trif}, \citenamefont {Stano},\ and\ \citenamefont
  {Loss}}]{Kloeffel2013}%
  \BibitemOpen
  \bibfield  {author} {\bibinfo {author} {\bibfnamefont {C.}~\bibnamefont
  {Kloeffel}}, \bibinfo {author} {\bibfnamefont {M.}~\bibnamefont {Trif}},
  \bibinfo {author} {\bibfnamefont {P.}~\bibnamefont {Stano}}, \ and\ \bibinfo
  {author} {\bibfnamefont {D.}~\bibnamefont {Loss}},\ }\href {\doibase
  10.1103/PhysRevB.88.241405} {\bibfield  {journal} {\bibinfo  {journal}
  {Physical Review B}\ }\textbf {\bibinfo {volume} {88}} (\bibinfo {year}
  {2013}),\ 10.1103/PhysRevB.88.241405}\BibitemShut {NoStop}%
\bibitem [{\citenamefont {Cubaynes}\ \emph {et~al.}(2019)\citenamefont
  {Cubaynes}, \citenamefont {Delbecq}, \citenamefont {Dartiailh}, \citenamefont
  {Assouly}, \citenamefont {Desjardins}, \citenamefont {Contamin},
  \citenamefont {Bruhat}, \citenamefont {Leghtas}, \citenamefont {Mallet},
  \citenamefont {Cottet},\ and\ \citenamefont {Kontos}}]{Cubaynes2019}%
  \BibitemOpen
  \bibfield  {author} {\bibinfo {author} {\bibfnamefont {T.}~\bibnamefont
  {Cubaynes}}, \bibinfo {author} {\bibfnamefont {M.~R.}\ \bibnamefont
  {Delbecq}}, \bibinfo {author} {\bibfnamefont {M.~C.}\ \bibnamefont
  {Dartiailh}}, \bibinfo {author} {\bibfnamefont {R.}~\bibnamefont {Assouly}},
  \bibinfo {author} {\bibfnamefont {M.~M.}\ \bibnamefont {Desjardins}},
  \bibinfo {author} {\bibfnamefont {L.~C.}\ \bibnamefont {Contamin}}, \bibinfo
  {author} {\bibfnamefont {L.~E.}\ \bibnamefont {Bruhat}}, \bibinfo {author}
  {\bibfnamefont {Z.}~\bibnamefont {Leghtas}}, \bibinfo {author} {\bibfnamefont
  {F.}~\bibnamefont {Mallet}}, \bibinfo {author} {\bibfnamefont
  {A.}~\bibnamefont {Cottet}}, \ and\ \bibinfo {author} {\bibfnamefont
  {T.}~\bibnamefont {Kontos}},\ }\href {http://arxiv.org/abs/1903.05229
  http://www.nature.com/articles/s41534-019-0169-4} {\bibfield  {journal}
  {\bibinfo  {journal} {npj Quantum Information}\ }\textbf {\bibinfo {volume}
  {5}},\ \bibinfo {pages} {47} (\bibinfo {year} {2019})}\BibitemShut {NoStop}%
\bibitem [{\citenamefont {Klinovaja}\ \emph {et~al.}(2012)\citenamefont
  {Klinovaja}, \citenamefont {Stano},\ and\ \citenamefont
  {Loss}}]{Klinovaja2012}%
  \BibitemOpen
  \bibfield  {author} {\bibinfo {author} {\bibfnamefont {J.}~\bibnamefont
  {Klinovaja}}, \bibinfo {author} {\bibfnamefont {P.}~\bibnamefont {Stano}}, \
  and\ \bibinfo {author} {\bibfnamefont {D.}~\bibnamefont {Loss}},\ }\href
  {https://link.aps.org/doi/10.1103/PhysRevLett.109.236801} {\bibfield
  {journal} {\bibinfo  {journal} {Physical Review Letters}\ }\textbf {\bibinfo
  {volume} {109}} (\bibinfo {year} {2012})}\BibitemShut {NoStop}%
\bibitem [{\citenamefont {Cottet}\ \emph {et~al.}(2017)\citenamefont {Cottet},
  \citenamefont {Dartiailh}, \citenamefont {Desjardins}, \citenamefont
  {Cubaynes}, \citenamefont {Contamin}, \citenamefont {Delbecq}, \citenamefont
  {Viennot}, \citenamefont {Bruhat}, \citenamefont {Dou{\c{c}}ot},\ and\
  \citenamefont {Kontos}}]{Cottet2017}%
  \BibitemOpen
  \bibfield  {author} {\bibinfo {author} {\bibfnamefont {A.}~\bibnamefont
  {Cottet}}, \bibinfo {author} {\bibfnamefont {M.~C.}\ \bibnamefont
  {Dartiailh}}, \bibinfo {author} {\bibfnamefont {M.~M.}\ \bibnamefont
  {Desjardins}}, \bibinfo {author} {\bibfnamefont {T.}~\bibnamefont
  {Cubaynes}}, \bibinfo {author} {\bibfnamefont {L.~C.}\ \bibnamefont
  {Contamin}}, \bibinfo {author} {\bibfnamefont {M.}~\bibnamefont {Delbecq}},
  \bibinfo {author} {\bibfnamefont {J.~J.}\ \bibnamefont {Viennot}}, \bibinfo
  {author} {\bibfnamefont {L.~E.}\ \bibnamefont {Bruhat}}, \bibinfo {author}
  {\bibfnamefont {B.}~\bibnamefont {Dou{\c{c}}ot}}, \ and\ \bibinfo {author}
  {\bibfnamefont {T.}~\bibnamefont {Kontos}},\ }\href {\doibase
  10.1088/1361-648X/aa7b4d} {\bibfield  {journal} {\bibinfo  {journal} {Journal
  of Physics Condensed Matter}\ }\textbf {\bibinfo {volume} {29}} (\bibinfo
  {year} {2017}),\ 10.1088/1361-648X/aa7b4d}\BibitemShut {NoStop}%
\bibitem [{\citenamefont {Cottet}\ and\ \citenamefont
  {Kontos}(2010)}]{Cottet2010}%
  \BibitemOpen
  \bibfield  {author} {\bibinfo {author} {\bibfnamefont {A.}~\bibnamefont
  {Cottet}}\ and\ \bibinfo {author} {\bibfnamefont {T.}~\bibnamefont
  {Kontos}},\ }\href@noop {} {\bibfield  {journal} {\bibinfo  {journal}
  {Physical Review Letters}\ }\textbf {\bibinfo {volume} {105}},\ \bibinfo
  {pages} {160502} (\bibinfo {year} {2010})}\BibitemShut {NoStop}%
\bibitem [{\citenamefont {Bulaev}\ and\ \citenamefont
  {Trauzettel}(2008)}]{Bulaev08}%
  \BibitemOpen
  \bibfield  {author} {\bibinfo {author} {\bibfnamefont {D.}~\bibnamefont
  {Bulaev}}\ and\ \bibinfo {author} {\bibfnamefont {D.}~\bibnamefont
  {Trauzettel}, \bibfnamefont {B.and~Loss}},\ }\href@noop {} {\bibfield
  {journal} {\bibinfo  {journal} {Physical Review B}\ }\textbf {\bibinfo
  {volume} {77}},\ \bibinfo {pages} {235301} (\bibinfo {year}
  {2008})}\BibitemShut {NoStop}%
\bibitem [{\citenamefont {Egger}\ and\ \citenamefont
  {Flensberg}(2012)}]{Egger12}%
  \BibitemOpen
  \bibfield  {author} {\bibinfo {author} {\bibfnamefont {R.}~\bibnamefont
  {Egger}}\ and\ \bibinfo {author} {\bibfnamefont {K.}~\bibnamefont
  {Flensberg}},\ }\href@noop {} {\bibfield  {journal} {\bibinfo  {journal}
  {Physical Review B}\ }\textbf {\bibinfo {volume} {85}},\ \bibinfo {pages}
  {235462} (\bibinfo {year} {2012})}\BibitemShut {NoStop}%
\bibitem [{\citenamefont {Laird}\ \emph {et~al.}(2015)\citenamefont {Laird},
  \citenamefont {Kuemmeth}, \citenamefont {Steele}, \citenamefont
  {Grove-Rasmussen}, \citenamefont {Nygård}, \citenamefont {Flensberg},\ and\
  \citenamefont {Kouwenhoven}}]{Laird2013}%
  \BibitemOpen
  \bibfield  {author} {\bibinfo {author} {\bibfnamefont {E.~A.}\ \bibnamefont
  {Laird}}, \bibinfo {author} {\bibfnamefont {F.}~\bibnamefont {Kuemmeth}},
  \bibinfo {author} {\bibfnamefont {G.~A.}\ \bibnamefont {Steele}}, \bibinfo
  {author} {\bibfnamefont {K.}~\bibnamefont {Grove-Rasmussen}}, \bibinfo
  {author} {\bibfnamefont {J.}~\bibnamefont {Nygård}}, \bibinfo {author}
  {\bibfnamefont {K.}~\bibnamefont {Flensberg}}, \ and\ \bibinfo {author}
  {\bibfnamefont {L.~P.}\ \bibnamefont {Kouwenhoven}},\ }\href {\doibase
  https://doi.org/10.1103/RevModPhys.87.703} {\bibfield  {journal} {\bibinfo
  {journal} {Reviews of Modern Physics}\ }\textbf {\bibinfo {volume} {87}},\
  \bibinfo {pages} {703} (\bibinfo {year} {2015})}\BibitemShut {NoStop}%
\end{thebibliography}
\end{document}